\documentclass[aps,prb,twocolumn,superscriptaddress]{revtex4}
\usepackage{graphicx}  
\usepackage{dcolumn}   
\usepackage{bm}        
\usepackage{amssymb}   
\usepackage{amsmath}
\usepackage{verbatim}
\usepackage{color}

\begin{document}

\title{Second harmonic generation from metallic arrays of rectangular holes}

\author{Sergio G. Rodrigo} \email{sergut@unizar.es}
\affiliation{Centro Universitario de la Defensa, Carretera de Huesca s/n, E-50090 Zaragoza, Spain} 
\affiliation{Instituto de Ciencia de Materiales de Arag\'on and
Departamento de F\'isica de la Materia Condensada,CSIC-Universidad
de Zaragoza, E-50009, Zaragoza, Spain}
\author{V. Laliena}
\affiliation{Escuela Universitaria Polit\'ecnica de La Almunia, Calle Mayor s/n, 50100 La Almunia, Spain}
\affiliation{Instituto de Ciencia de Materiales de Arag\'on and
Departamento de F\'isica de la Materia Condensada,CSIC-Universidad
de Zaragoza, E-50009, Zaragoza, Spain}
\author{L. Mart\'{i}n-Moreno}
\affiliation{Instituto de Ciencia de Materiales de Arag\'on and
Departamento de F\'isica de la Materia Condensada,CSIC-Universidad
de Zaragoza, E-50009, Zaragoza, Spain}

\date{\today}

\begin{abstract}
The generation process of second harmonic radiation from holes periodically arranged on a metal surface is investigated. Three main modulating factors affecting the far-field distribution and the transmission efficiency are identified: the near-field distribution at the wavelength of the driving source (fundamental harmonic), how second harmonic light couples to the diffraction orders of the lattice and its propagation properties inside the holes. It is shown that light generated at second harmonic can excite electromagnetic modes otherwise unaccessible in the linear regime under normal incidence illumination, a singularity of second harmonic fields that affects the radiation process. For instance, the least decaying transversal electric $\text{TE}_{0,1}$ mode accessible to the external beam is able to generate a superposition of high order modes ($\text{TE}_{1,1}$ and $\text{TM}_{1,1}$) at second harmonic. It is demonstrated that the emission of second harmonic radiation is only allowed along off-normal paths precisely due to that symmetry. In this work, two different regimes are studied in the context of extraordinary optical transmission, where enhanced linear transmission either occurs through localized electromagnetic modes or is aided by surface plasmon polaritons. While localized resonances in metallic hole arrays have been previously investigated, the role played by surface plasmons in second harmonic generation has not been addressed so far. In general, good agreement is found between our calculations (based on the finite difference time domain method) and the experimental results on localized resonances, even though no free fitting parameters were used in describing the materials. It is found that second harmonic emission is strongly modulated by enhanced fields at the fundamental wavelength (either localized or surface plasmon modes) on the glass-metal interface. This is so in the transmission side but also in reflection, where emission can only be explained by an efficient tunneling of second harmonic photons through the holes from the output to the input side. Finally, the existence of a \textit{dark} surface plasmon polariton at the fundamental field is identified through a non-invasive method for the first time, by analyzing the efficiency and far-field pattern distribution in transmission at second harmonic.
\end{abstract}


\maketitle 

\section{Introduction}\label{intro}
Second harmonic generation (SHG) is a non-linear process that creates a single photon at $\lambda/2$ through the interaction of two photons of wavelength $\lambda$~\cite{FrankenPRL61}. Since its discovery, SHG has become a current toolkit in optics with potential applications in several areas: from research in biological imaging~\cite{CampagnolaNatureBio03,PantazisPNAS10} to recent advances in quantum information through parametric down-conversion [inverse second harmonic (SH) process]~\cite{WalbornPhysRep10}. The SH fields originate from the bulk and the surface of many substances. In centrosymmetric materials SHG is electric-dipole forbidden in the bulk, so only the first high order leading terms (electric-quadrupole, magnetic-dipole...) contribute to SHG~\cite{BloembergenPhysRev68}. On the other hand, SH fields may generate at the surface where the inversion symmetry is broken~\cite{SipePRB80,CorviPRB86,GuyotSionnestPRB88,MizrahiJOptSocAmB88}. The bulk contribution can not be neglected in flat metal surfaces~\cite{XiangPRB09} and may be greatly enhanced in nanostructures due to the presence of large field gradients at the surface~\cite{ScaloraPRA10,BachelierPRB10,BenedettiOptExp11}.

Initially, most of the experiments that investigated SHG from metals focused in flat surfaces, but the advance in nanofabrication techniques and optical characterization at the nanoscale~\cite{ZayatsPhysicsReports04,Maier07,Bozhevolnyi09,AtwaterNatureMater10,Novotny12,HoonAdvFuncMat12} has turned the attention to SH effects in metallic nanostructures. For instance, some theoretical predictions on SHG from spherical nano-particles, developed over more than two decades~\cite{AgarwalSolidStateCommun82,HuaPRB86,OestlingZPhysD93,DadapPRL99,DadapJOptSocAmB04,CiraciPRB12}, have been experimentally tested at single particle level only recently~\cite{ButetNanoLett10}. Nano optics know-how opens the possibility to future SHG-based applications for optical characterization~\cite{ButetNanoLett12,RodrigoPRL13,ButetOptExp13,HaoJournalChemPhys02,BozhevolnyiPRB02,HaykBookChap12}.
\begin{figure}[thb!]
\centering\includegraphics[width=0.6\columnwidth]{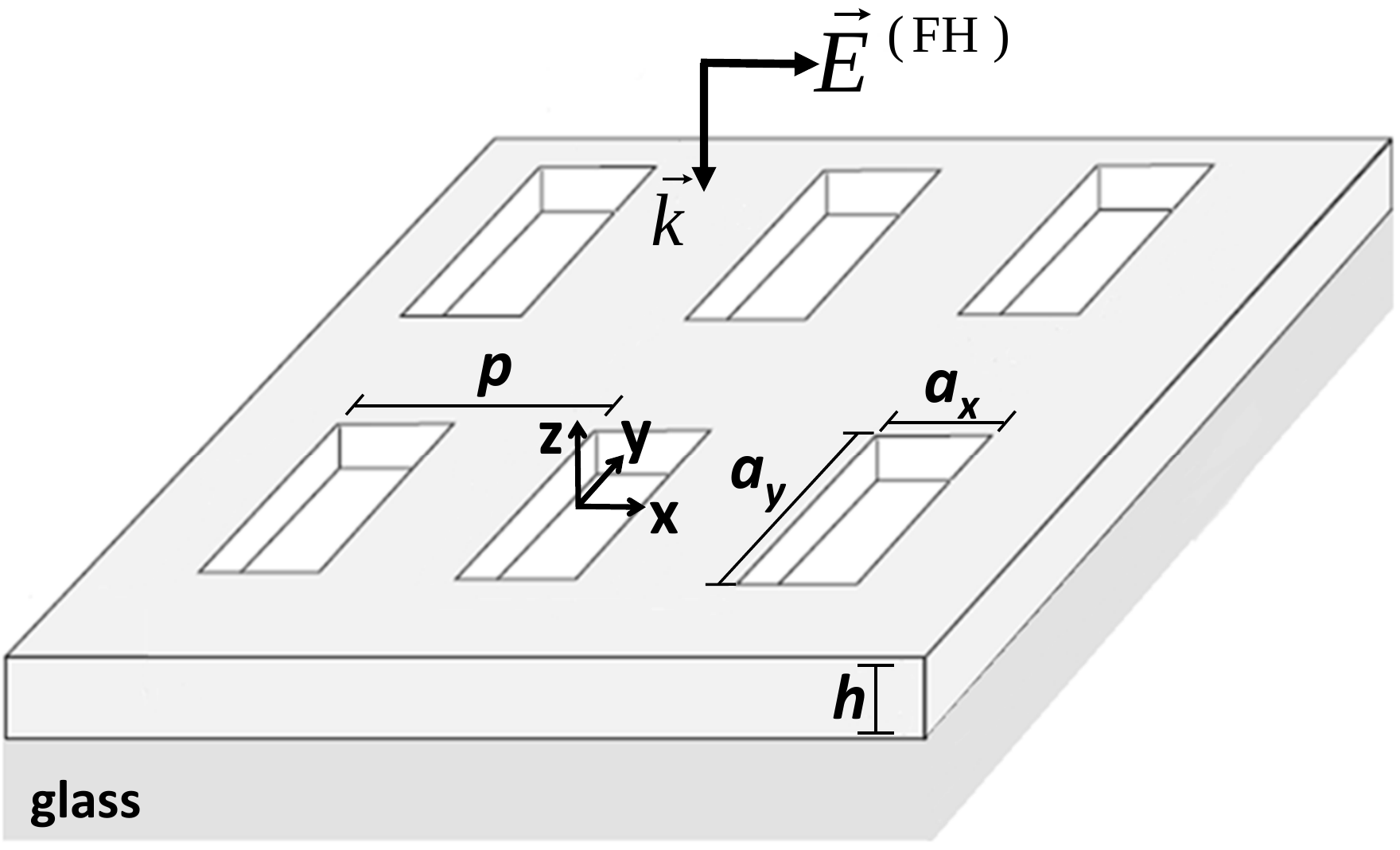} \caption{Schematics of an array of rectangular holes deposited on glass. The illumination at the fundamental harmonic (air side) is polarized along the x-axis, normally incident to the surface.} \label{fig0}
\end{figure}

The SHG process is very weak in a flat metal surface, so only high-intensity lasers provide enough output. Nanostructured metal films locally enhance the intensity of the incident field, which might be useful to obtain SHG at less demanding laser powers. This is the case of an array of holes drilled in a metal film, a nanostructured system which has been widely investigated in the linear regime especially since the discovery of extraordinary optical transmission (EOT)~\cite{EbbesenNature98}. Taking advantage of the strong electromagnetic (EM) fields found in holey films, several attempts have been conducted to exploit their nonlinear response for harmonic generation~\cite{AirolaJOptA05,WurtzPRL06,NieuwstadtPRL06,XuOptExp07,EftekhariIEEE08,EvgeniaPRB08,SchaichPRB08,WurtzLaserPhotRev08,SchonOptLett10,VincentiOptExp11,RodrigoPRB11,WangSciRep13}. 

In particular we are interested in the seminal work by Nieuwstadt and coworkers~\cite{NieuwstadtPRL06} on SHG emission from two dimensional rectangular hole arrays (RHAs) carved on gold films (see Fig.~\ref{fig0}). They found SH enhancement due to localized modes occurring close to the cutoff wavelength of the fundamental harmonic (FH) field, $\lambda_c$. That enhancement was explained in terms of slow EM modes localized in the holes at FH. But the conclusion has been challenged by the same authors~\cite{PrangsmaNJP10}, and recent experiments have shown that the time delay at FH is similar for different aspect ratios~\cite{WangSciRep13}. 

So many questions are still unsolved about the nonlinear SHG response in such important plasmonic platform. Can the experimental results be explained? Where does SH emission come from? Which are the characteristics of the EM modes excited at SH inside the holes? How is SH radiation spatially distributed at the far-field? Also, although it is known that Surface Plasmon Polaritons (SPPs) are at the origin on EOT, their possible influence on SHG has not been investigated yet, up to the best of our knowledge. In this work, we try to answer these and further questions.

The paper is organized as follows: in Sec.~\ref{model} we describe the theoretical approach, which is based on the finite difference time domain (FDTD) method~\cite{Taflove05}. In Sec.~\ref{genprop} we describe the system geometry and the source of FH fields. We also discuss the main mechanisms governing SHG from RHAs. In particular, we describe three main factors that modulate the nonlinear response. In Sec.~\ref{sim} we analyze the coupling of SH light with the diffraction orders of the lattice, which is demonstrated to be controlled both by the symmetry of the SH field and the lattice period. We explain why only off-normal non-evanescent diffraction orders are allowed for SH emission (forbidden through the $z$-direction), under normal incidence illumination. Assuming that SH emission occurs at both transmission and reflection half-spaces, the FH near-field (Sec.~\ref{FieldDis}) and the propagation properties of SH light inside the holes (Sec.~\ref{Abs}) determine the balance of SHG between these regions. In Sec.~\ref{resFH} we extend our previous analysis on SPPs and throughly analyze the optical response at SH triggered by localized resonances at FH (we compare our numerical results with the experimental measurements of Ref.~\cite{NieuwstadtPRL06}). We end up with the conclusions in Sec.~\ref{conclusion}.

\section{Theoretical approach}\label{model}
We use the FDTD method for calculations. The linear response of gold is described by the Drude-Lorentz model presented in Ref.~\cite{HaoChemPhysLett07}, which provides suitable functional fit to experimental data valid in the visible range~\cite{JohnsonPRB72}. Regarding the outer boundaries, we simulate infinite periodic hole arrays thorough the application of appropriate (Bloch) conditions at the boundaries of the unit cell ($x-$ and $y-$ directions) and imposing ``uniaxial perfect matched layers'' at surfaces parallel to the metal film through the $z$-direction~\cite{Taflove05}. To calculate the optical response at  SH frequency we follow a perturbative approach, for which the fundamental field is not affected by the SH field (non-depletion approximation). This is an excellent approximation given that the radiated intensity at FH is roughly ten orders of magnitude larger than at SH.  We solved simultaneously the system of coupled equations governing the propagation of the FH and SH fields. The equations for the FH field obviously coincide with the linear Maxwell equations when no SH field is present, while the SH field obeys the Maxwell equations with sources determined by the FH field through the second order polarization vector, $\textbf{P}^{(\text{SH})}$. 

We used the following surfacelike model for the induced polarization at SH frequency:
\begin{eqnarray}
\textbf{P}_n^{(\text{SH})} &=& \left[\chi^{(2)}_{nnn}
\vert E_n^{(\text{FH})}\vert^2+\chi^{(2)}_{ntt}
\vert E_t^{(\text{FH})}\vert^2 \right] \textbf{n} \nonumber \\
\textbf{P}_t^{(\text{SH})}&=& 2\chi^{(2)}_{tnt}
E_n^{(\text{FH})}\textbf{E}_t^{(\text{FH})} \label{eq1}
\label{Eq1}
\end{eqnarray}
where $n$ and $t$ stand for normal and tangential to the surface respectively, and $\chi_{ijk}$ are the non-vanishing components of an effective second-order susceptibility tensor. The FH electric field is taken at the metal surface, and from it $\textbf{P}^{(\text{SH})}$ is calculated at the same location.

Given the uncertainty in measuring $\chi^{(2)}$, which quite often can even differ from one sample to another, we assumed that the nonlinear response of gold is weakly dispersive around the frequency range investigated and for definiteness we take the (internal field) effective second-order susceptibility from Ref.~\cite{XiangPRB09}, that is: $\chi^{(2)}_{nnn}= 250.0$, $\chi^{(2)}_{tnt}= 3.6$, and $\chi^{(2)}_{ntt}= 1.0 $ in units of $3.27~10^{-15}$ cm/V.  In this way the effective nonlinear susceptibility contains both surface and bulk contributions. Note that Eq.~\ref{eq1} includes the fundamental bulk contribution, $\gamma \nabla [\textbf{E}^{(\text{FH})} . \textbf{E}^{(\text{FH})}]$ (being $\gamma$ a material parameter), which is treated in an effective manner. For homogeneous and isotropic materials there is another volume source, which has been neglected here, the only one that can be measured separately from the surface, $\delta [\textbf{E}^{(\text{FH})} . \nabla] \textbf{E}^{(\text{FH})}$ (again $\delta$ depends on the material). It has been found that $\delta$ is so weak that the separable bulk contribution plays thus a minor role in SHG from flat metal surfaces~\cite{XiangPRB09}, and it seems to be also negligible in the case of nanospheres~\cite{BachelierPRB10,CaprettiPRB14}.

Note however all bulk contributions can be cast into a surfacelike model~\cite{GuyotSionnestPRB88}, an approach which is valid in the limit of vanishing nonlocal response, as recently demonstrated from first principles by Ciraci et al. within the framework of the hydrodynamic model~\cite{CiraciPRB12}.

In FDTD, every spatial location in the system can be visualized as a cube with the electric field components pointing along the edges and the magnetic field components being normal to the faces~\cite{Taflove05}. The interface between two different materials is composed of adjacent faces, the electric field components lying on the interface. When a face rests on a metal surface the electric field is computed with the piece linear recursive convolution method~\cite{LuebbersIEEE92}. Outside the metal, the electric field is updated as corresponds to a lossless dielectric. This  method  applies  both  to  SH  and  FH  fields  so the hole surfaces are treated in the same way. Every metal face belongs to a given FDTD cell and has associated a single value of $\textbf{P}^{(\text{SH})}$. The tangential electric field in Eq.~\ref{eq1} is obtained from the average of the electric fields located at the face edges. The normal component of the electric field in Eq.~\ref{eq1} is obtained from the average of the electric fields located at the edges perpendicular to that face, inside the metal. 

We intentionally work at fixed wavelength and change the geometrical parameters defining the array, using realistic values similar to that of experiments~\cite{NieuwstadtPRL06}. In this way, we separate the spectral contributions to SHG arising from the geometry to those arising from the spectral dependence of the nonlinear polarizability (which is not well known). 

In many circumstances, Eq.~\ref{Eq1} can be simplified by neglecting the contributions to SH from $\chi^{(2)}_{tnt}$ and $\chi^{(2)}_{ntt}$, provided SH radiation keeps unaltered under this approximation. This is the case here and up to 80$\%$ of the total radiated photons at SH originate from $\chi^{(2)}_{nnn}$, as we will show later on. The normal surface component preserves the full symmetry  of the SH field and provides qualitative results compared to the full theoretical treatment (including all the tensor components of Eq.~\ref{eq1}). Under this approximation, the far-field power at SH is  proportional to $[\chi^{(2)}_{nnn}]^2$, which becomes the only free parameter to describe SHG in the systems under study. We believe this simplification will make future comparisons between experiments and theory easier.

Finally, let us mention that we checked our code against analytical results obtained for flat metallic surfaces~\cite{DonnellNJP05} obtaining accurate results for a numerical mesh size of $5$~nm (not shown), which has been used in the following RHA calculations.

\section{Far-field emission at SH: general properties}\label{genprop}
Holey metal films display a complex optical linear response, which is mainly controlled by the geometrical parameters of  the array, the optical properties of the metal and the refractive index of the surrounding medium~\cite{GarciaVidalRevModPhys09}. Here, we describe the main differences between the linear and the nonlinear response, the most important: FH and SH fields have opposite parity symmetry. At the same wavelength, light inside the holes can couple to waveguide modes of different symmetry whether it originates from an external source (laser beam) or it is nonlinearly generated at SH. These EM modes are unaccessible in the linear regime at normal incidence, while they are allowed for oblique incidence. In  the  last  case  however,  such  a  mode  cannot  be  isolated  from  the EM modes with opposite symmetry, those which can be linearly excited at normal incidence. Therefore symmetry determines the ability of the local SH fields to couple with the propagating diffraction orders of the lattice. For instance, emission of SH light is only allowed for off-normal propagation at normal incidence. In addition, the radiation pattern and intensity at FH is essential to understand SHG, but also the propagation properties of the waveguide modes excited at SH inside the holes. 

The intensity of SH radiation depends on both the material properties (linear dielectric constant and $\chi^{(2)}$) and the geometry of the system in a complicated manner. In what follows, we describe three main factors that modulate the nonlinear response: i) coupling of SH light with the lattice diffraction orders (controlled by the symmetry of SH fields); ii) local field distribution at FH; and iii) the propagation properties of SH light inside the holes.

In the calculations, the FH beam is a truncated plane-wave at $\lambda_{\text{FH}}=830$~nm. The source illuminates the system at normal incidence from the air side with the electric field pointing along the x-axis. The intensity used, $0.1~\text{MW/cm}^2$, has been estimated from the linear power measurement reported in Fig.~3 of Ref.~\cite{NieuwstadtPRL06}. For a detailed explanation about the plane-wave source conditions and the method to calculate scattering coefficients see Ref.~\cite{RodrigoPRB11}. The whole system is on a glass substrate ($n_{\text{glass}}=1.5$).

A note about geometry: roughly, the optical response of a RHA at $\lambda_{\text{FH}}$ has localized character for configurations with $n_{\text{glass}} \, p << \lambda_c\approx \lambda_{\text{FH}}$. On the other hand, the scattering properties are dominated by the coupling to surfaces modes at larger periods. In general, some degree of hybridization between localized modes and surface waves always exists~\cite{MaryPRB07}. In this section, the period is varied and the hole shape and film thickness are fixed ($a_x=a_y=280$~nm ; $h=160$~nm). These geometrical parameters are chosen so that we can explore both optical regimes keeping the incident wavelength unchanged. 

\subsection{Symmetry of SH fields: far-field coupling}\label{sim}
\begin{figure}[htbp] 
\centering\includegraphics[width=\columnwidth]{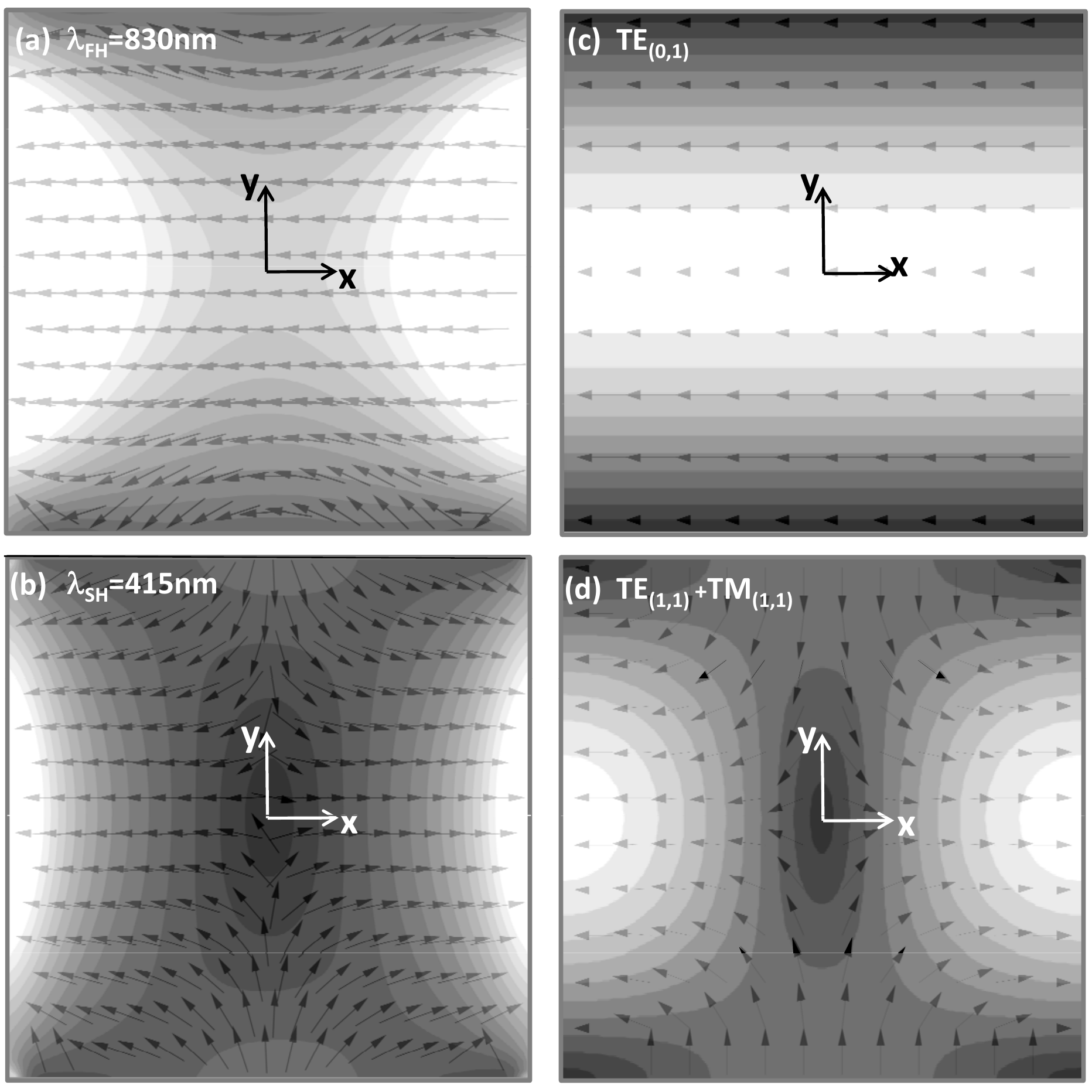} \caption{(a) Numerical calculation of the electric field components lying on a x-y plane inside a hole for $p=500$~nm, at FH ($\lambda_{\text{FH}}=$830~nm) in gold. Lateral dimensions: $a_x=a_y=280$~nm. Film thickness: $h=160~\text{nm}$. The plane of observation is situated at $z=-125$~nm.  The amplitude is also represented, superimposed to the vector field map [gray scale: white (maximum) and black (minimum)]. (b) The same but for SH ($\lambda_{\text{SH}}=415$~nm). (c) Analytical calculation of the $\text{TE}_{0,1}$ mode electric field, for a perfect electric conducting infinite waveguide [same lateral dimensions as in (a)]. (d) The same but for a superposition of the $\text{TE}_{1,1}$ and $\text{TM}_{1,1}$ modes. Only those electric field components lying on the x-y plane are represented for the $\text{TM}_{1,1}$ mode.} \label{fig2}
\end{figure}

Figure~\ref{fig2} shows the computed electric field on a plane placed at $z=-125$~nm inside a single hole of an array with $p=500$~nm: for (a) $\lambda_{\text{FH}}=830$~nm and (b) $\lambda_{\text{SH}}=415$~nm. The amplitude is superimposed to the vector field map.  

The electric field distribution of the $\text{TE}_{0,1}$ mode for a perfect electric conducting (PEC) infinite waveguide (with the considered cross section) was analytically calculated and shown in Fig.~\ref{fig2}(c). Usually, the identification between EM modes in real and PEC metals is justified in EOT~\cite{GarciaVidalRevModPhys09}. The finite conductivity of real metals explains the deviations observed in Fig.~\ref{fig2} between panels (a) and (c). At first approximation, it translates to an effective enlargement of the hole size due to the EM field penetration inside the metal. Furthermore, a redistribution of the field inside the holes occurs because of the plasmonic nature of such EM fields (see Ref.~\cite{GordonOptExp05}). In any case, the symmetry of the electric field pattern for PEC is the same as the one for gold, which allows one to conclude from direct inspection that the FH field corresponds to the $\text{TE}_{0,1}$ mode. The cutoff wavelengths for gold and PEC infinite waveguides are $\lambda_c^{\text{TE}_{0,1}}=712$~nm and $\lambda_c^{\text{TE}_{0,1}}=2\,a_y=560$~nm, respectively.

The symmetry properties of the waveguide modes excited can be characterized by the action of the operator $\Pi_j$, which provides the parity symmetry of the electric field along the j-direction over the orthogonal plane that crosses the center of the hole. 

The parity along the y-direction, $\Pi_y$, is identical for the modes excited at FH and SH. The y-component of the incident electric field is zero, so it does not impose any constraint through that direction.

The parity along the x-direction, $\Pi_x$, is different for the FH and SH fields. The x-component (y-component) of the $\text{TE}_{0,1}$ mode has even (odd) parity, that is, $\Pi_x \, E_x^{(\text{FH})}(x,y)=  + E_x^{(\text{FH})}(-x,y)$ and $\Pi_x \, E_y^{(\text{FH})}(x,y)=  - E_y^{(\text{FH})}(-x,y)$, for both gold and PEC. The incident field, a linearly polarized plane-wave, only can excite EM modes with that symmetry at normal incidence. The electric field at SH has opposite symmetry through the action of $\Pi_x$ [see Fig.~\ref{fig2}(b)]. We observe that $\Pi_x E_x^{(\text{SH})}(x,y)= -E_x^{(\text{SH})}(-x,y)$ and $\Pi_x E_y^{(\text{SH})}(x,y)= +E_y^{(\text{SH})}(-x,y)$. 

Therefore, the $\text{TE}_{0,1}$ mode is forbidden at SH for normal incidence illumination. The vector field pattern at SH for gold [Fig.~\ref{fig2}(b)] results from the excitation of higher order EM modes of the infinite waveguide, which can be demonstrated by the same procedure used for the FH field. Analyzing the dispersion relation of the PEC rectangular waveguide, we find that $\text{TE}_{1,1}$ and $\text{TM}_{1,1}$ modes are the next least decaying modes (cutoff wavelengths: $\lambda_c^{\text{TM}_{1,1}}=\lambda_c^{\text{TE}_{1,1}}=\frac{2 a_x a_y}{\sqrt{a_x^2+ a_y^2}}=396$~nm). These modes have the right symmetry, so they might be excited at SH.  However none of them individually leads to the field profile expected from Fig.~\ref{fig2}(b). Instead, a superposition of the $\text{TE}_{1,1}$ and $\text{TM}_{1,1}$ modes provides a near-field pattern inside the holes compatible with the SH field, as it is shown in Fig.~\ref{fig2}(d). The relative amplitude and phase of each mode has been adjusted to reproduce the numerical result. A different aspect ratio or hole shape would end up in a different balance between the corresponding waveguide modes~\cite{RodrigoOptExp10,SchonOptLett10}. The cutoff of both modes in the gold waveguide redshifts compared to the PEC case as expected, $\lambda_c^{\text{TM}_{1,1}}=508$~nm and $\lambda_c^{\text{TE}_{1,1}}=585$~nm. Therefore, the propagation of SH fields through the holes is perfectly possible at $\lambda_{\text{SH}}=415$~nm, being the amplitude of SH fields only affected by the absorption in the metal.

The change of symmetry can be readily explained. The mirror symmetry of SH fields results from the properties of the second order polarization vector. According to the approximated expression $\textbf{P}^{(\text{SH})}\approx\chi^{(2)}_{nnn}\vert E_n^{(\text{FH})}\vert^2 \textbf{n}$ for Eq.~\ref{eq1}, the direction of $\textbf{P}^{(\text{SH})}$ at a given point is approximately determined by the unitary vector normal to the surface at that location.  By definition, \textbf{n} is positive at the left side wall of the hole ($x=-a_x/2$), and negative at the right side wall ($x=+a_x/2$).  The symmetry of the charge density at SH switches from odd to even because $\vert E_n^{(\text{FH})}\vert^2$ is equal at both sides [Fig.~\ref{fig2}(a) and Fig.~\ref{fig2}(c)]. 

Symmetry is crucial in the SHG process and determines the ability of the SH field to couple with the propagating diffraction orders in the substrate and the incident half-space. In a periodic structure only  propagation through those directions given by the Bragg's condition are allowed at normal incidence, set by the parallel to the surface reciprocal lattice vector, $\textbf{G}_{i,j}=\frac{2 \pi}{p}(i,j)$ for a square lattice ($i$ and $j$ are integers). Using the dispersion relation of light in the substrate we calculate the angle $\theta_{i,j}$ for the $(i,j)$-order in transmission with respect to the vertical direction, expressed as function of the fundamental wavelength: 
\begin{eqnarray}
\sin{\left ( \theta_{ij} \right )}=  \dfrac{\sqrt{i^2+j^2}}{n_{\text{glass}}}  \left ( \dfrac{\lambda_{\text{FH}}}{2\,p}\right )
\label{Eq2}
\end{eqnarray}
Propagating modes are then associated to absolute values of $\sin{\left ( \theta_{ij} \right )}$ equal or less than unity, yielding evanescent modes otherwise. In the reflection side we can calculate the corresponding angles by taking $n_{\text{glass}}=1$. We can distinguish among the different diffracted orders with FDTD by projecting onto diffracted modes in each dielectric half-space. The basic idea consists in finding a way to isolate the current that each wave-vector of the reciprocal lattice carries, as a function of both the wavelength and the polarization state (for further details see Ref.~\cite{RodrigoTESIS} and references therein). 

The situation is different in the case of SHG. Note that the EM field of the 0th diffraction order is constant on a given $x-y$ plane, so it has the same parity symmetry that of the incident field at normal incidence. As a result, SH and 0th diffraction order fields have opposite parity symmetry, so the overlap between both EM modes is zero. Therefore, there is no SH radiation in a half-space with all the diffraction orders being evanescent except the 0th diffraction order. In our calculations, SH emission in transmission at the 0th diffraction order, $J_{\text{SH}}^T(0,0)$, is sixteen orders of magnitude (at the level of noise due to numerical round off) less intense than through other directions, for all periods. The same occurs in the reflection region, confirming that our FDTD implementation fully respects the symmetry of SH fields. Similar arguments based on symmetry considerations explain why specular radiation is forbidden for regular arrangements of nanoparticles~\cite{McMahonPRB06}. Moreover SH photons with $\textbf{G}_{i,j}^{(\text{SH})}=0$ are forbidden under the normal incidence condition, so only the evanescent waves on the lattice at FH (SPPs, non-propagating near-field scattered by the holes...) can create SH fields obeying momentum conservation~\cite{BloembergenJOSA80}.
\begin{figure}[thb!]
\centering\includegraphics[width=0.7\columnwidth]{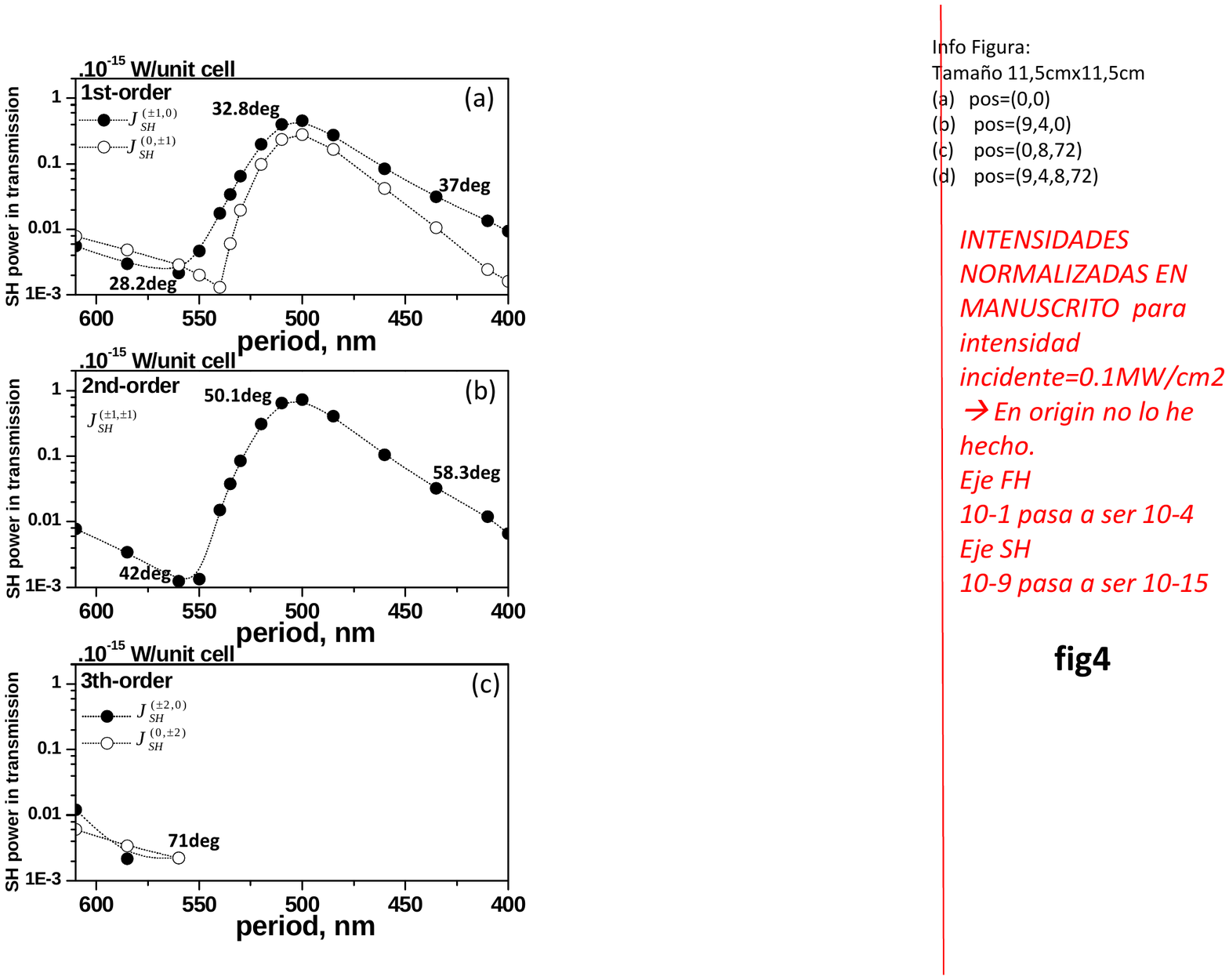} \caption{(a), (b) and (c): SH power in transmission per unit cell  through the allowed diffraction orders in glass, as a function of the period. The rest of geometrical parameters are: $h=160~\text{nm}$ and $a_x=a_y=280$~nm.} \label{fig4}
\end{figure}

Interestingly, the different diffraction orders that contribute to the same $\theta_{i,j}$ do not need to have the same intensity. As example, we define two SH powers in transmission per unit cell for the 1st diffraction order,  which are represented in Fig.~\ref{fig4}(a) as a function of the period. The parallel component, $J_{\text{SH}}^T(\pm1,0)=J_{\text{SH}}^T(+1,0)+J_{\text{SH}}^T(-1,0)$, accounts for the SH fields radiated on two planes defined by vectors $\textbf{G}_{\pm1,0}$, which are parallel to the electric field of the FH source.  The perpendicular component $J_{\text{SH}}^T(0,\pm1)=J_{\text{SH}}^T(0,+1)+J_{\text{SH}}^T(0,-1)$ represents the same but for the planes given by the corresponding diffraction orders, which are perpendicular to the electric field of the FH source in this case. Parallel and perpendicular components at SH are different, a result that will be explained later on. The 2nd diffraction order $J_{\text{SH}}^T(\pm1,\pm1)$,  the sum of the four possible combinations of the  $(\pm1,\pm1)$ diffraction orders, is represented in Fig.~\ref{fig4}(b). In this case, because the reciprocal lattice vectors are parallel to the diagonals of the unit cell each order carries the same SH intensity. Finally, the results for the 3th diffraction order are shown in panel  Fig.~\ref{fig4}(c).  The 3th diffraction order becomes evanescent for periods shorter than $553$~nm, as predicted by Eq.~\ref{Eq2}. Note that the angle of emission increases as the period size decreases, for a given diffraction order ($\theta$ is shown for a few periods in Fig.~\ref{fig4}). 

The emission of SHG into the transmission region splits in different angular contributions, which can be controlled by the period size. On the other hand, the profile of total SH emission in transmission as a function of the period is characterized by a dip followed by a peak, both features related to the excitation at FH of SPPs (on the glass-metal interface). They are determined by the field intensity and distribution at FH. This very influent factor on SHG is discussed next.

\subsection{Field distribution at FH: local source of SH fields}\label{FieldDis}
\begin{figure}[thb!]
\centering\includegraphics[width=0.7\columnwidth]{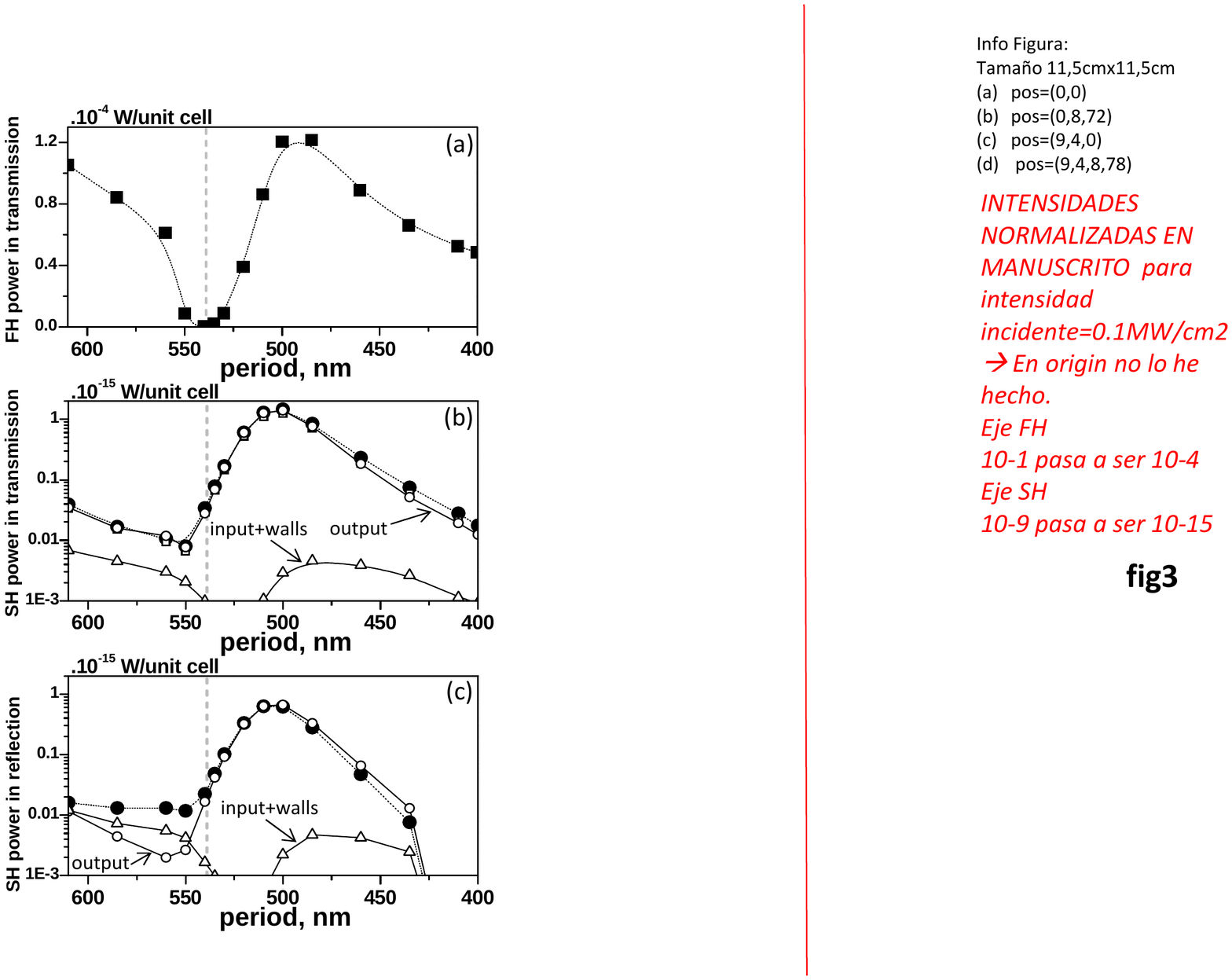}
\caption{(a) Transmitted FH power per unit cell as a function of the period for $\lambda_{\text{FH}}=$830~nm. (b) SH power in transmission per unit cell as a function of the period at $\lambda_{\text{SH}}=$415~nm. Solid circular symbols: full calculation; empty square symbols: an approximation taking $\chi^{(2)}_{nnn}\neq0$ (neglecting both $\chi^{(2)}_{ntt}$ and $\chi^{(2)}_{tnt}$); empty circular symbols: an approximation taking $\chi^{(2)}=0$ everywhere, except on the output surface; and empty triangular symbols: approximation taking $\chi^{(2)}=0$ everywhere, except on the input surface and walls. (c) Same as in (b) but for the SH intensity emitted in the reflection region. The vertical lines indicate the period for which $\lambda_{\text{FH}}$ coincides with the EOT minimum. The rest of geometrical parameters are: $h=160~\text{nm}$ and $a_x=a_y=280$~nm.}
\label{fig3}
\end{figure}

In Fig.~\ref{fig3}(a) the FH power transmitted per unit cell, $J_{\text{FH}}^T$, is calculated for several periods ranging from $610$~nm to $400$~nm. The dip at $p\approx540$~nm corresponds to an EOT minimum (indicated with a vertical line, as a guide to the eye). The EOT maximum at FH is reached for $p\approx 500$~nm. The corresponding SH  powers radiated in transmission and reflection per unit cell, $J_{\text{SH}}^T$ and $J_{\text{SH}}^R$, are shown in Fig.~\ref{fig3}(b) and Fig.~\ref{fig3}(c) with solid circular symbols. Figure~\ref{fig3}(b) also shows with empty squares the SH power emitted in transmission calculated by neglecting  $\chi^{(2)}_{ntt}$ and $\chi^{(2)}_{tnt}$ contributions (but keeping $\chi^{(2)}_{nnn}\neq0$) showing that is an excellent approximation, as advanced in the introduction.

Second harmonic generation fundamentally depends on the specific details (phase and intensity) of $\textbf{E}^{(\text{FH})}$ on the metal surface, unlike for example two-photon luminescence that essentially depends on the local intensity~\cite{BeermannPhysStatusSolidiC05}.  Let us analyze the EM modes supported by the investigated structures, which are ultimately related to the features of the linear transmission spectrum in Fig.~\ref{fig3}(a).

\begin{figure*}[htbp] 
\centering\includegraphics[width=0.7\textwidth]{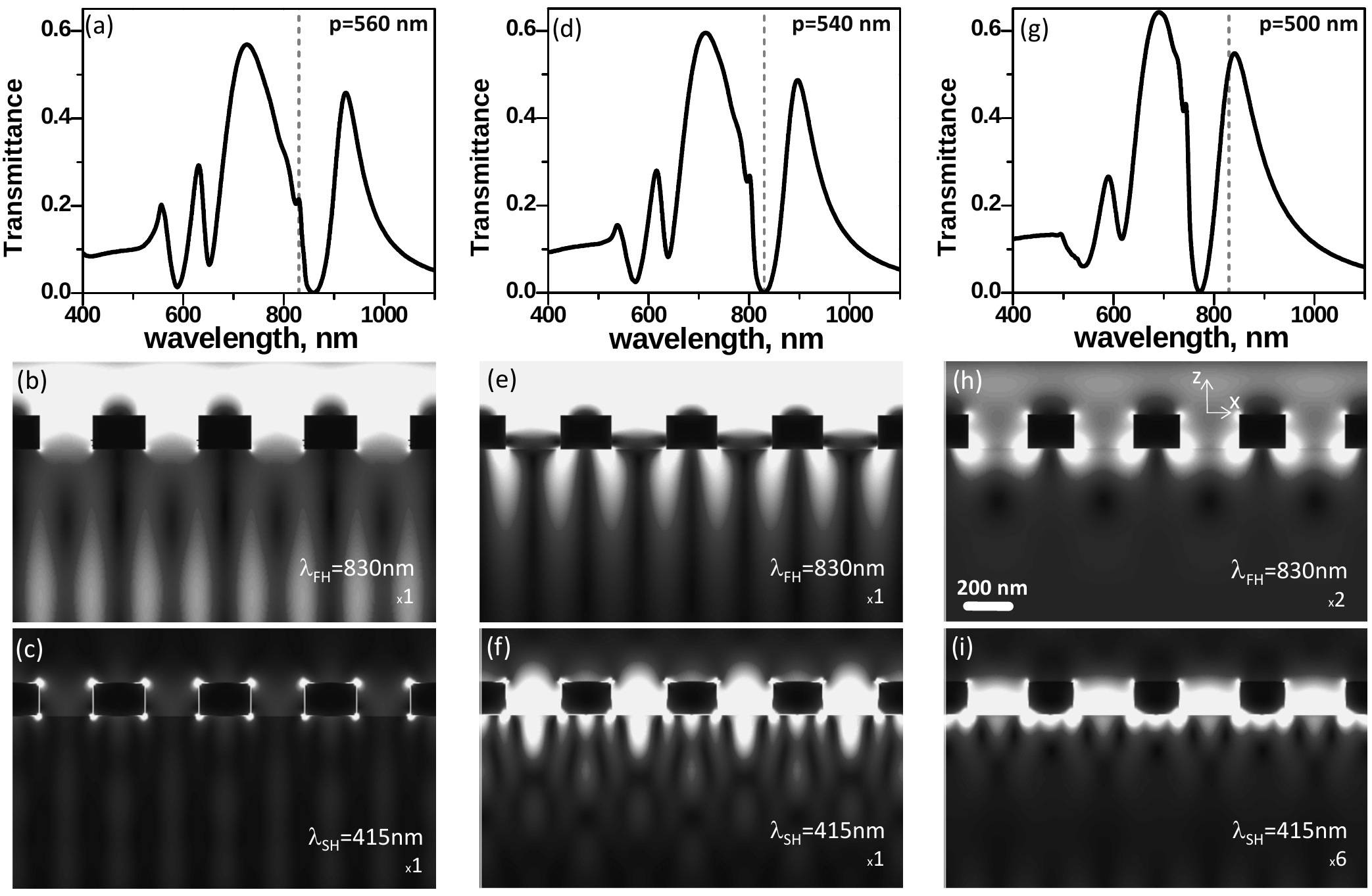} \caption{For a RHA in gold with period $p=560~\text{nm}$ illuminated at normal incidence from the input side (see Fig~\ref{fig0}): (a) linear transmittance, (b) and (c) electric field amplitude at FH ($\lambda_{\text{FH}}=830$~nm) and SH for several unit cells along the $x$-direction, calculated on a plane placed at $y=0$. (d)-(f) Same calculation but for $p=540~\text{nm}$. (g)-(i) Same calculation but for $p=500~\text{nm}$. Gray scale: white (maximum) and black (minimum). Relative scale shown in bottom right corners. Same geometrical parameters as in Fig.~\ref{fig4} and Fig.~\ref{fig3}.} \label{fig1}
\end{figure*}

The linear transmission spectra for periods $p=560$~nm, $p=540$~nm and $p=500$~nm are shown in panels (a), (d) and (g) of Fig.~\ref{fig1}. The vertical lines depict the wavelength of the external source $\lambda_{\text{FH}}$. Three EOT peaks can be distinguished in each figure, resulting from the excitation of surface EM modes of the corrugated structure~\cite{GarciaVidalRevModPhys09}. Each surface mode has associated a full EOT feature, characterized by the typical profile of a Fano resonance~\cite{GenetOptCommun03}. At the EOT minimum the SPP of the holey film is hardly affected by the presence of holes~\cite{RodrigoPRB08}, so we can identify every peak in Fig.~\ref{fig1} with the help of the flat surface dispersion relation for SPPs, $k_{\text{SPP}}$~\cite{BarnesNature03}. The frequency at which a SPP can be excited at normal incidence, $\lambda_{\text{SPP}}$, is given by the condition of momentum conservation at the surface and can be approximately calculated by folding $k_{\text{SPP}}$ into the first Brillouin's zone, i.e., $k_{\text{SPP}}=\vert \textbf{G}_{i,j} \vert$ (where $\textbf{G}_{i,j}$ is a reciprocal lattice vector). For example, the EOT minima in Fig.~\ref{fig1}(d) ($p=540$~nm) are located at wavelengths: $830$~nm, $640$~nm and $574$~nm. These energies correspond to three different bounded EM modes. The EOT peak at the near infrared is due to the $(\pm1,0)\text{-SPP}$ of the glass-metal interface, being $\lambda_{\text{SPP}}=844$~nm. The prominent feature at visible is mainly due to the glass-metal $(\pm1,\pm1)\text{-SPP}$, with $\lambda_{\text{SPP}}=636$~nm. Slightly blue-shifted, we can also see a less intense and more narrow peak which corresponds to the air-metal $(\pm1,0)\text{-SPP}$, with $\lambda_{\text{SPP}}=578$~nm. 
It is reasonable to think that the field distribution of SPPs supported by a RHA determines which surface in the system is the main source of SH radiation. Our formalism allow us to switch off the generation of SH fields from a given surface (by forcing $\chi^{(2)}=0$ at that surface). This allows us to obtain insight into the predominant SH process. We have calculated $J_{\text{SH}}^T$ and $J_{\text{SH}}^R$ as all SHG would be exclusively generated from three different regions separately: transmission side (output), reflection side (input) and hole walls. The output and input contributions include SHG from the edges and corners of the holes. The radiation in transmission at SH originating  from the output interface, $J_{\text{SH}}^{T,out.}$, is shown in Fig.~\ref{fig3}(b) with empty circular symbols, while the remaining SH radiation (input+walls) is shown with triangles. The corresponding results for the reflection region are shown in Fig.~\ref{fig3}(c). Note that the superposition principle for Maxwell's equations only applies to vector fields, so in general $J_{\text{SH}}^{T} \neq J_{\text{SH}}^{T,out.}+J_{\text{SH}}^{T,in.}+J_{\text{SH}}^{T,walls}$ (the same in the reflection half-space). If one of the contributions take over the rest the equality approximately holds.

In transmission, the glass-metal $(\pm1,\pm1)\text{-SPP}$ is responsible for radiation at SH in transmission from $p=610$~nm to $p=550$~nm, given that $J_{\text{SH}}^{T} \approx J_{\text{SH}}^{T, out.}$. However, the contribution to SH from walls and input surface is not negligible [see triangular symbols in Fig.~\ref{fig3}(b)], which is coherent with near-field at SH [Fig.~\ref{fig1}(c)]. For periods ranging from $p=540$~nm to $p=400$~nm  $J_{\text{SH}}^{T} \approx J_{\text{SH}}^{T, out.} > J_{\text{SH}}^{T, in.}+J_{\text{SH}}^{T, walls}$, so generation at the output region accounts for most of the SH emission in transmission. This result is also coherent with the SH near field maps shown in Fig.~\ref{fig1}. On the output surface, the glass-metal $(\pm1,0)\text{-SPP}$ develops within that period range.  This SPP has a clear fingerprint in the FH near-field at the glass-metal interface, seen in Fig.~\ref{fig1}(e) and Fig.~\ref{fig1}(h) (see relative scale). At the EOT minimum the near-field is intense enough to generate strong local SH fields [Fig.~\ref{fig1}(f)], which are comparable to those generated  at maximum transmission (only six times more intense) [Fig.~\ref{fig1}(i)]. The optical response at SH in transmission has then a straightforward explanation: overall the generation of SH radiation in transmission is controlled by enhanced fields at FH on the glass-metal interface because of the excitation of surface waves bound to that surface, as expected for an asymmetric dielectric configuration~\cite{krishnanOptCommun01}.

In the reflection region, the radiation process at SH presents three different regimes [see Fig.~\ref{fig3}(c)].  From $610$~nm to $550$~nm, $J_{\text{SH}}^{R, out.} \leq J_{\text{SH}}^{R, in.}+J_{\text{SH}}^{R, walls}$. The interpretation is that both walls and input side contributions to SH are important in reflection. For that period range, the system has access to the $(\pm1,\pm1)\text{-SPP}$ at FH, which is bound to the glass-metal surface. In the input side, the FH field is enhanced at the holes [Fig.~\ref{fig1}(b)]. This field is characterized by a combination of localized and surface modes. In fact, the near-field within that period range is affected by the optical response of a single hole as reported in Ref.~\cite{MaryPRB07}, given the close proximity between the FH wavelength and the hole cutoff ($\approx 712$~nm).  Therefore, the FH field is distributed at both sides of the metal layer and inside the holes, yielding a more complex optical response at SH, specially in reflection. From the period at which the EOT minimum occurs for the chosen incident wavelength, up to $p=435$~nm we find that $J_{\text{SH}}^R \approx J_{\text{SH}}^{R, out.}$. The FH field is asymmetrically distributed and concentrates on the output surface as it corresponds to the glass-metal $(\pm1,0)\text{-SPP}$. To explain this behavior we need to understand how SH light propagates inside the holes, which is discussed in the following paragraphs. Finally, for $p < \lambda_\text{\text{SH}}$  all off-normal diffraction orders are evanescent in air, so SH radiation in reflection is zero within the round off precision in our numerical calculations.

\subsection{Propagation inside the holes at SH: light absorption}\label{Abs}
In Sec.~\ref{sim} we advanced that propagation inside the holes is only limited by light absorption in the metal, given that $\lambda_{\text{SH}} < \lambda_c^{\text{TM}_{1,1}} < \lambda_c^{\text{TE}_{1,1}}$, for the investigated hole dimensions. Within the period range where SHG is caused by the glass-metal $(\pm1,0)\text{-SPP}$, the SH fields created at the glass-metal interface can go through the holes and be emitted into the reflection region, which explains that most of the SH emission in reflection originates at the output surface. Moreover, $J_{\text{SH}}^T$ and $J_{\text{SH}}^R$ intensities have the same order of magnitude [compare Fig.~\ref{fig3}(b) with Fig.~\ref{fig3}(c)], except for the shortest periods where $J_{\text{SH}}^R = 0$. The amount of SH light in reflection is not negligible, so light absorption is not limiting the SH emission process.
\begin{figure}[thb!]
\centering\includegraphics[width=0.7\columnwidth]{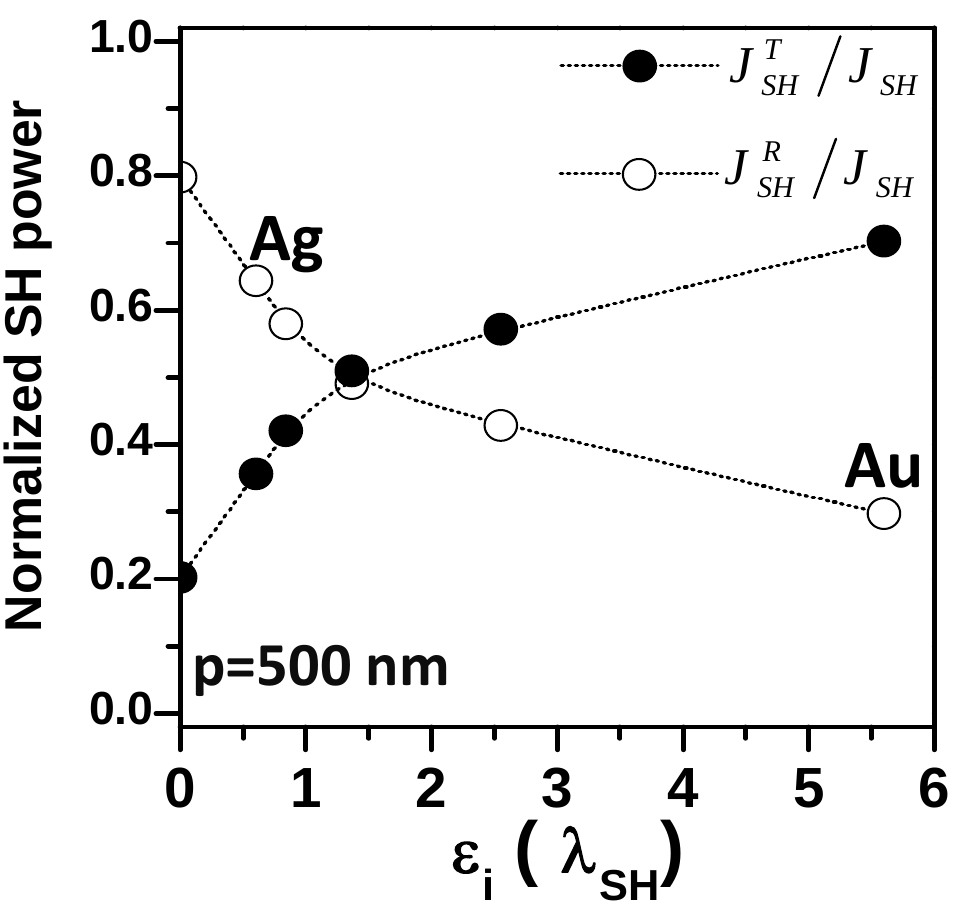}
\caption{For $p=500$~nm, SH power in transmission and in reflection normalized to total SH emission, as a function of the imaginary part of the dielectric constant of gold, silver and a series of hypothetical metals. The real part of the dielectric constant is essentially that of Ag in the series. The rest of geometrical parameters are: $h=160~\text{nm}$ and $a_x=a_y=280$~nm.}
\label{fig6}
\end{figure}
To further investigate the consequences of light absorption inside the holes we compare gold with a series of hypothetical metals, for which the real part of the dielectric constant is essentially that of silver but the imaginary part, $\varepsilon_i$, is modified. The value of $\chi^{(2)}$ in these simulations is the one used here for gold. The SH powers in transmission and in reflection over total SHG are shown in Fig.~\ref{fig6}, as a function of the imaginary part of the dielectric constant at SH. We have chosen $p=500$~nm, which illustrates the optical response of the system when the glass-metal $(\pm1,0)\text{-SPP}$ is excited. The skin depth, the cutoff wavelengths of the waveguide modes at SH, and the near field pattern at FH are practically the same in all the series. We have also found that $J_{\text{SH}}^T \approx J_{\text{SH}}^{T, out.}$ and $J_{\text{SH}}^R \approx J_{\text{SH}}^{R, out.}$ for all $\varepsilon_i$ considered (not shown). In the lossless case $J_{\text{SH}}^R > J_{\text{SH}}^T$. The difference between both of them reduces with increasing absorption inside the holes. Eventually the inequality in reversed, and $J_{\text{SH}}^T > J_{\text{SH}}^R$, providing a clear evidence of the critical role played by metal absorption.   

\section{Resonances at FH}\label{resFH}
\subsection{SPP related effects}\label{SPP}
In this section, the linear and the nonlinear response around the glass-metal $(\pm1,0)$-SPP is analyzed in more detail. We define two optical properties to characterize emission: SH efficiency in transmission and the change of polarization of the first diffraction order.
 
\begin{figure}[thb!]
\centering\includegraphics[width=\columnwidth]{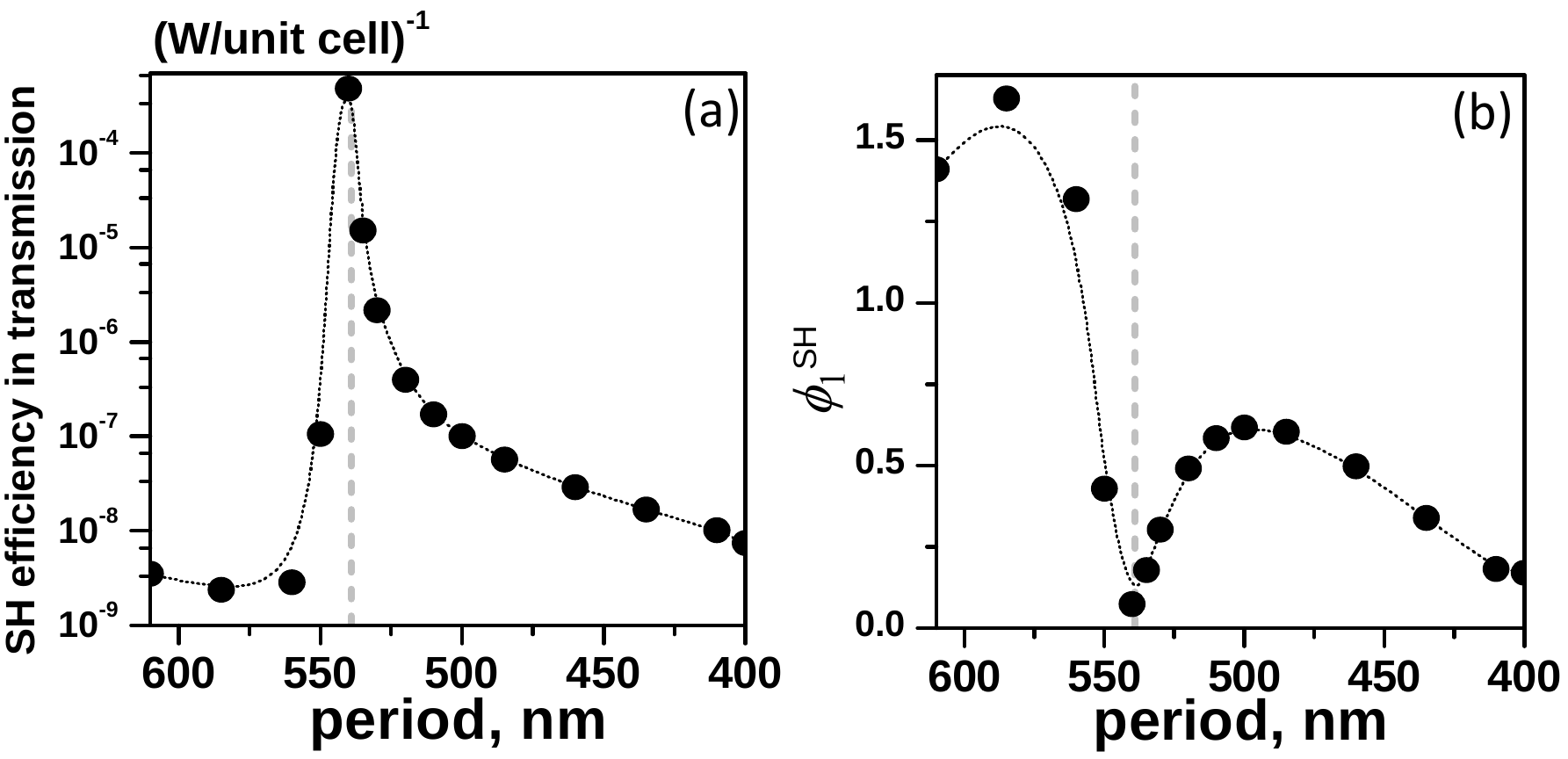}
\caption{(a) SH efficiency as a function of the period. (b) The change of polarization calculated for the 1st diffraction order, $\phi^{\text{SH}}_1$. For definitions, see the main text. The vertical lines indicate the period for which $\lambda_{\text{FH}}$ coincides with the EOT minimum. The rest of geometrical parameters are: $h=160~\text{nm}$ and $a_x=a_y=280$~nm.}
\label{fig7}
\end{figure}

In Fig.~\ref{fig7}(a) we show the results for SH efficiency in transmission, defined as:
\begin{eqnarray}
\alpha= J_{\text{SH}}^T/(J_{\text{FH}}^T)^2
\label{Eq3}
\end{eqnarray}
This coefficient is independent of the illumination intensity at FH. The vertical line, indicating the period for which the wavelength of the considered incident beam $\lambda_{\text{FH}}$ corresponds to an EOT minimum at FH, coincides with the  maximum value of $\alpha$.
Efficiency at the EOT minimum is five orders of magnitude larger than the minimum value obtained among the periods investigated. The spatial distribution of the FH field becomes again the key point. The optical properties of a SPP in a holey metal film are different at different wavelengths. For convenience, we distinguish between the response at the EOT minimum and the response at the transmission maximum using different labels in each case. At the EOT minimum the SPP of the holey surface remains ``unperturbed'', showing the same optical response than that of a SPP on a flat metal surface, as explained in Sec.~\ref{FieldDis}. Therefore, the influence of holes can be ignored assuming that the EM field at their centers is negligible. A mode like this is weakly coupled to the far-field, so we call it dark-SPP.  The SPP of the corrugated structure is highly ``perturbed''  at the EOT maximum, so its dispersion relation deviates from that of a flat metal surface. The holes scatter light and we call it bright-SPP. Efficiency reaches the highest value at the EOT minimum because the dark-SPP is weakly coupled to the far-field in transmission at FH [Fig.~\ref{fig3}(a)], while its near-field at the output surface creates enough SH photons in the transmission region [Fig.~\ref{fig3}(b)].

Figure~\ref{fig4}(a) shows that the far-field pattern of SHG is not symmetrically distributed, thus the so-called parallel and perpendicular components of SH radiation for the first diffraction order are different [$J_{\text{SH}}^T(\pm1,0) \neq J_{\text{SH}}^T(0,\pm1)$]. We define the change of polarization for the 1st diffraction order as:
\begin{eqnarray}
\phi^{\text{SH}}_1=J_{\text{SH}}^T(0,\pm1)/J_{\text{SH}}^T(\pm1,0)
\label{Eq4}
\end{eqnarray}
shown in Fig.~\ref{fig7}(b). The parallel component dominates the far-field emission for linearly generated light at the same wavelength of SH [$J_{\text{FH}}^T(0,\pm1)/J_{\text{FH}}^T(\pm1,0)\approx 0$]. We could expect $\phi^{\text{SH}}_1\approx 0$, however $\phi^{\text{SH}}_1$ reaches values even higher than the unity. We observe that $J_{\text{SH}}^T(\pm1,0)$ dominates for short periods, while for large periods the balance reverses. In between, a sudden change in the SHG spatial distribution exists, at the period for which $\lambda_{\text{FH}}$ approximately matches the EOT minimum. 

As explained, the coupling with surface modes at FH is very weak for short periods. The radiative and even the non-propagating field polarizaton follow that of the incident field, which is the typical response found in isolated holes~\cite{DeLeonPRL12}. As the period increases, the bright-SPP starts to participate in the generation process. Its wavevector is directed along the x-direction, however this plasmon mode is efficiently scattered by the holes and may provoke, for every interaction with them, a strong depolarization of the evanescent FH field thus increasing $\phi^{\text{SH}}_1$. The same argument can be sustained for the largest periods investigated, for which the glass-metal $(\pm1,\pm1)$-SPP is excited at FH. A singular behavior occurs at the EOT minimum, at the dark-SPP. This mode hardly scatters light at FH, so SH radiation can be only generated along the direction of its wavevector, i.e., the x axis.

Similar results are expected in experimental setups by tuning $\lambda_{\text{FH}}$, for a fixed period. In that situation, a measure of the change of polarization might be useful for surface assays, where the balance between the radiated intensity through different directions would determine the quality of a given sample. Surface defects, shape imperfections, debris on the surface, and whatever other experimental realization far from a perfect hole array would have an impairing impact on $\phi^{\text{SH}}_1$. In addition, the EOT minimum (which in simulations appears as a sharp and narrow dip) is quite sensitive to sample imperfections and the size of the system~\cite{BravoAbadNatPhys06}. Therefore, SH efficiency would provide additional information on high quality structure characterization~\cite{HoonAdvFuncMat12}. Deviations from the expected $\alpha$ profile [Fig.~\ref{fig7}(a)], like broadening, would indicate a mis-alignment of the FH beam, structure imperfections generated during the fabrication process or the presence of chemical products on the surface. 

\subsection{Localized resonances and related effects}\label{loc}
The influence of localized resonances in SHG from RHAs was first studied in Ref.~\cite{NieuwstadtPRL06}.  In this work enhanced SH emission occurring close to the cutoff wavelength of the FH field was reported. That enhancement was explained in terms of slow EM modes localized in the holes at FH. But the conclusion has been challenged by the same authors~\cite{PrangsmaNJP10}, and recent experiments have shown that the time delay at FH is similar for different aspect ratios~\cite{WangSciRep13}. 

In that experimental work several hole arrays were investigated, each consisting of $20\text{x}20$ rectangular holes milled in a square lattice. The system was deposited on a glass substrate. The film thickness, period and hole area were kept fixed at $h=160$~nm, $p=410$~nm and $S=3.4 \text{x} 10^4~\text{nm}^2$ respectively.  The period was precisely chosen to avoid hybridization between localized modes and surface waves at FH~\cite{MaryPRB07}.  The different arrays were designed with holes of different dimensions, characterized by the aspect ratio, $\text{AR}$. The system was illuminated at normal incidence from the air side, being $\lambda_{\text{FH}}=830$~nm and with FH peak powers ranging from approximately $1.0~\text{mW}$ to $50.0~\text{mW}$ (intensities from $1.5~\text{KW/cm}^2$ to $0.075~\text{MW/cm}^2$). We use the same parameters than in the experimental work, but our x-y axes are rotated 90~degrees with respect to Ref.~\cite{NieuwstadtPRL06}. With our choice, $\text{AR}=a_y/a_x$, $a_x=\sqrt{\text{S/AR}}$ and $a_y=\sqrt{\text{AR\,x\,S}}$. Like in previous calculations, the FH beam is a truncated plane-wave at $\lambda_{\text{FH}}=830$~nm. The FH source illuminates the system at normal incidence from the air side with the electric field pointing along the x-axis and delivers $0.1~\text{MW/cm}^2$.
\begin{figure}[thb!]
\centering\includegraphics[width=\columnwidth]{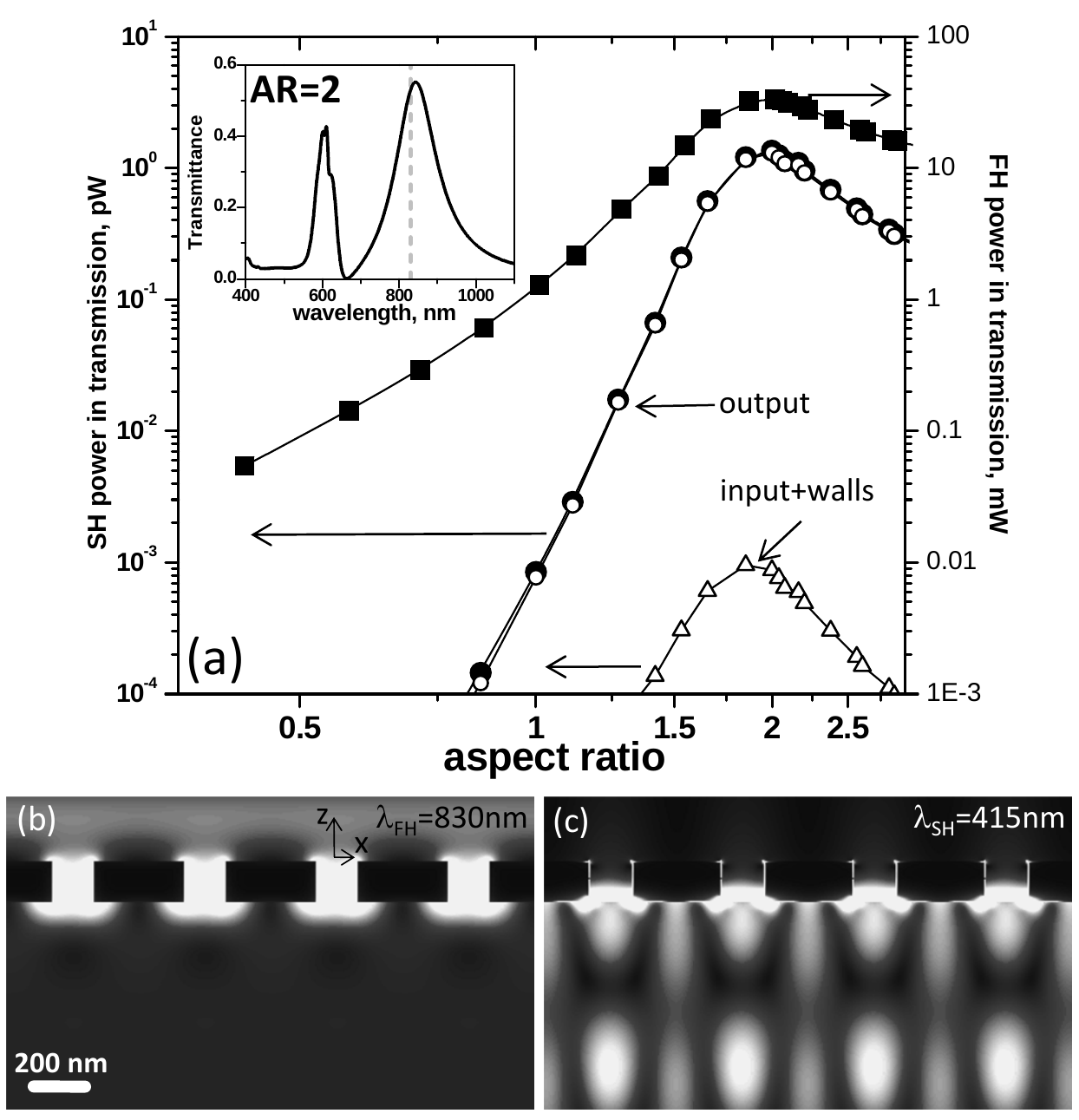} \caption{(a) FH and SH power emitted in transmission from a surface covering $20\text{x}20$ unit cells shown with full square and circular symbols respectively, as a function of the aspect ratio ($\text{AR}=a_y/a_x$). The empty circular symbols shows the approximation taking $\chi^{(2)}=0$ everywhere, except in the output surface. The empty triangular symbols is the approximation taking $\chi^{(2)}=0$ everywhere, except in the input surface and walls. The inset shows the linear transmission spectrum calculated for $\text{AR}=2$. The vertical line indicates the wavelength $\lambda_{\text{FH}}=$830~nm, used for calculations in the main panel. (b) and (c) Near field maps for AR=2. The electric field amplitude at FH ($\lambda_{\text{FH}}=830$~nm) and SH for several unit cells along the $x$-direction, are calculated on a plane placed at $y=0$. Gray scale: white (maximum) and black (minimum). Same scale of Fig.~\ref{fig1}(h) and Fig.~\ref{fig1}(i) for FH and SH, respectively. The FH incident field is polarized along the x-axis. The geometrical parameters are: $p=410$~nm and $h=160$~nm.} \label{fig5}
\end{figure}

We present in Fig.~\ref{fig5}(a) the computed power emitted in transmission at FH (square symbols) and SH (full circular symbols),  as a function of the aspect ratio. From the power per unit cell calculated with FDTD, we can directly compare our results with the experimental ones, taking into account that the samples covered $20\text{x}20$ unit cells. The simulations reproduce the general trend of experiments [Fig.~3 in Ref.~\cite{NieuwstadtPRL06}]. For instance, both linear and nonlinear transmissions peak at AR~$\approx 2$ with similar SHG power values. In contrast, the measured linear transmission is flatter as a function of AR, and our result on SHG is characterized by a broader peak compared with the experimental case. In any case, the main features of SHG from RHAs are captured by our numerical implementation, which is even more notorious given the sensitivity of $\chi^{(2)}$ to sample imperfections or the presence of chemical byproducts on the surface and that no free fitting parameters have been used. Therefore, the numerical approach developed is suitable for describing the nonlinear behavior of such metallic nanostructures.
\begin{figure}[thb!]
\centering\includegraphics[width=\columnwidth]{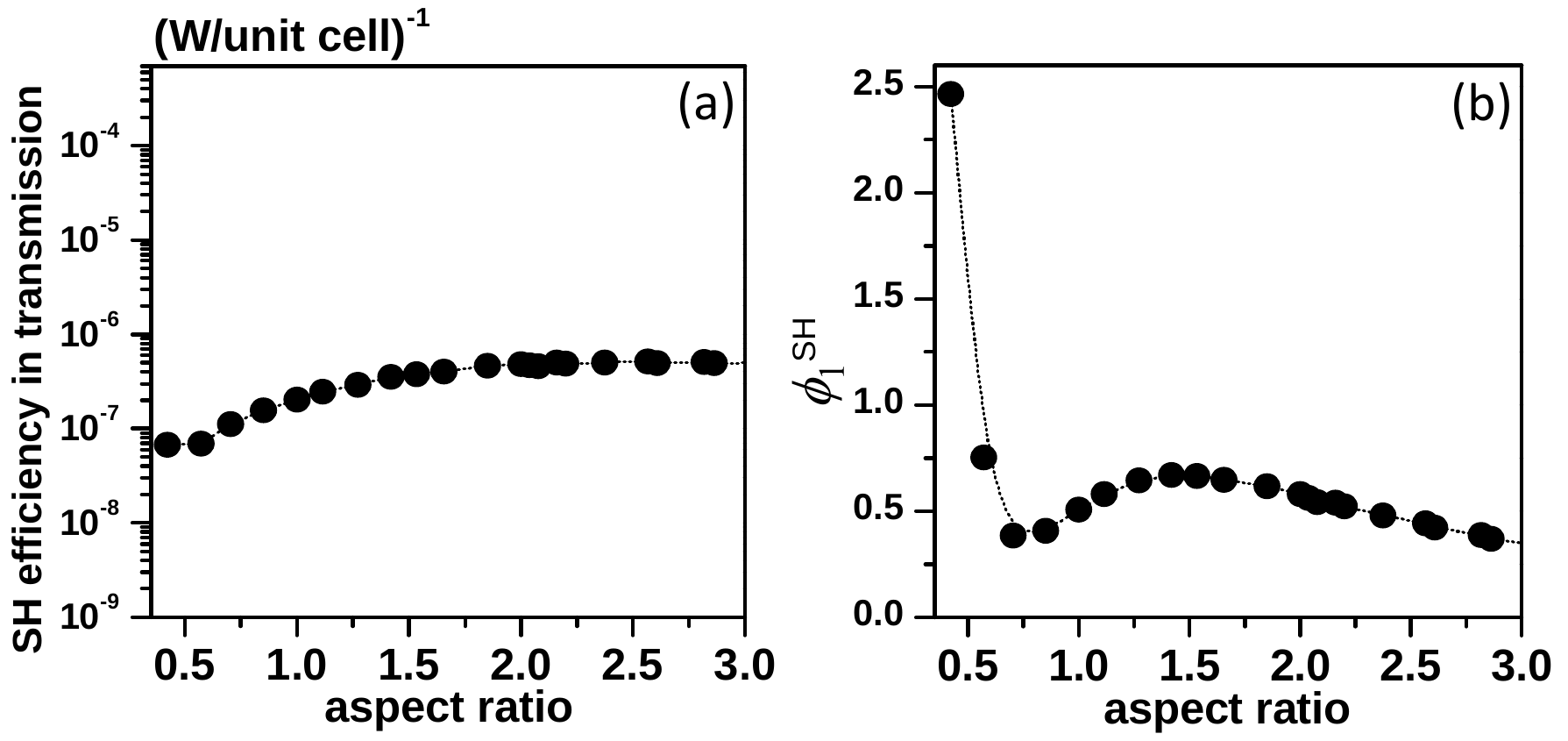}
\caption{(a) $\alpha$ and (b) $\phi^{\text{SH}}_1$ as a function of the aspect ratio. For definitions, see the main text. Same parameters of Fig.~\ref{fig5}.}
\label{fig8}
\end{figure}  
This system can be analyzed at the light of the physical mechanisms presented up to here. First, linear and non-linear peak locations coincide like in Fig.~\ref{fig3}, where enhanced SH emission occurs at the maximum of linear transmission. In Fig.~\ref{fig5}(a), the SHG peak is due to a localized resonance~\cite{KoerkampPRL04,GarciaVidalPRL05,CarreteroPalaciosPRB12} excited at the FH wavelength (for an example, see the inset of Fig.~\ref{fig5}(a), which represents the FH transmission spectrum at AR=2). At resonance, EM fields are bound to the structure for time scales long enough to generate EM field accumulation at FH. Resonances at SH might not be discarded, however we have not observed any related effect either for surface waves or in localized resonances.  Second, SH emission in the reflection region (air side) is forbidden for the chosen period, as expected from Eq.~\ref{Eq2}. Third, the FH field is more intense at the output surface than at the input surface because of the presence of the dielectric. A mode tends to concentrate its EM energy in regions of high refractive index~\cite{Joannopoulos08} as shown in Fig.~\ref{fig5}(b) for AR=2, producing both strong SH fields at the output [Fig.~\ref{fig5}(c)] and high SH emission in transmission. Again, we can analyze SH radiation from three different regions separately: transmission side (output), reflection side (input) and hole walls. The radiation in transmission at SH originating  from the output interface, $J_{\text{SH}}^{T,out.}$, is shown in Fig.~\ref{fig5}(a) with empty circular symbols, while the remaining SH radiation (input+walls) is shown with triangles. As expected for analyzing the SH near-field, $J_{\text{SH}}^T \approx J_{\text{SH}}^{T, out.}$ for all aspect ratios.   

Finally, the corresponding SH efficiency and $\phi^{\text{SH}}_1$ is shown in Fig.~\ref{fig8}, panels (a) and (b), as a function of the aspect ratio. Efficiency does not show the abrupt change observed in Fig.~\ref{fig7}(a). The change of polarization displays two different regimes, below and beyond AR~$\approx 1$. For AR~$<1$ the short side of the hole is perpendicular to the incident electric field. In this case we expect a strong change in the polarization of light (from x to y polarization) even at FH. For AR~$>1$ the long side of the rectangles is now perpendicular to the incident electric field, so there is less depolarization. Having in mind the results of Sec.~\ref{genprop}, we realize that the period is of utmost relevance compared to other geometrical parameters in SHG from RHAs.  
 
\section{Conclusions}\label{conclusion}
In conclusion, we have investigated SH radiation in periodic arrays of rectangular holes drilled in a metal film. We have conducted FDTD calculations to compare with existing experimental works, where the effect of localized resonances in SHG was studied. Our simulations, which do not contain any fitting parameter, are able to capture the general trends of experimental data, which demonstrates that FDTD simulations are suitable for describing the nonlinear behavior of such metallic structures. 
We have investigated the role played by surface plasmon polaritons on SHG from RHAs, which has not been previously discussed, to the best of our knowledge. We have demonstrated that the generation of SH radiation in transmission is mainly controlled by enhanced fields (either localized or SPP modes) at FH on the glass-metal interface. In the reflection side, the same modes are responsible of emission thanks to the efficient tunneling of SH photons through the holes.  Because SHG fundamentally depends on the specific details of these EM fields, we have shown how SHG can be a non-invasive method for probing the FH near field. For the first time, the excitation of a SPP dark-mode in a metallic planar structure is identified by analyzing the efficiency and far-field pattern distribution at SH.  We have explained all these findings through a subtle physical mechanism. The SH near field, induced by the FH currents, has opposite parity symmetry from that of the FH field and provides access to EM modes that cannot be excited by the fundamental field at normal incidence. Ultimately, it is the character of such modes (absorption, cutoff wavelength, overlap with the lattice diffraction orders) what determines whether SH light can be emitted to the far-field or not. As expected, the emission of SH light is only allowed for off-normal propagation at normal incidence illumination. We have seen that the far-field distribution and efficiency of SHG into the transmission region strongly depends on the period size being hardly affected by the aspect ratio of holes. We believe our findings will be useful as tool for checking the quality of any kind of holey metal system.

\vspace{0.2cm}

\textbf{Acknowledgments}\\
We acknowledge support from the Spanish Ministry of Science and Innovation under projects MAT2011-28581-C02, and CSD2007-046-Nanolight.es.


\end{document}